# The Role of Public Health in the Fight Against Cancer: Awareness, Prevention, and Early Detection

**Narges Ramezani**

Department of Biology, Damghan Branch, Islamic Azad University, Damghan, Iran.

**Erfan Mohammadi**

Faculty of Entrepreneurship, University of Tehran, Tehran, Iran.

**Abstract**

Cancer, as a complex and devastating sickness, poses a great public health challenge on a global scale. Its far-attaining impact necessitates a devoted branch of medicine referred to as oncology, which makes a specialty in the prevention, diagnosis, and treatment of cancer. In this paper, we aim to offer a complete evaluation of the oncology sector, delving into its rich history, numerous forms of most cancers, diagnostic methods, and treatment alternatives. By exploring recent advances in oncology, which include precision remedy, immunotherapy, and the integration of era, we shed mild on the promising traits of the subject. However, it is miles essential to know the continual challenges that we face, such as the high costs related to remedy and the emergence of drug resistance. Despite these challenges, the final goal of oncology remains unwavering - to provide exceptional feasible outcomes for sufferers of most cancers, using both healing and palliative remedy strategies. As our knowledge of this complicated ailment continues to adapt, we must prioritize prevention and early detection and deal with disparities in access to care. By fostering collaboration and operating collectively, we can seriously improve the lives of tens of millions of individuals affected by cancer around the arena.

**Introduction**

Poor physical and intellectual fitness, such as conditions like tension, despair, and misery, can significantly increase the threat of growing various oncological sicknesses (Xu et al., 2020; Zhang et al., 2020; Gong et al., 2021; Zhang et al., 2020; Chen et al., 2021). Therefore, it is vital to prioritize bodily and intellectual well-being (Bai et al., 2023) to reduce the probability of developing such illnesses (Ferlay et al., 2010). The oncology records can be traced back to when cancer was first defined with the aid of the historical Egyptians. The Greek physician Hippocrates also wrote about cancer in the 5th century BCE. However, significant progress in oncology was no longer made until the nineteenth century. In 1838, the German pathologist Johannes Muller proposed that most cancers turned into because of the out-of-control boom of cells (Jones & Baylin, 2007). Subsequent studies also proved this concept, forming the inspiration for our contemporary knowledge of cancer. Cancer is one of the main reasons for the demise of internationally, and its prevalence is expected to continue rising in the coming years.

Oncology, as a department of medicine, is dedicated to the prevention, analysis, and treatment of most cancers. It is a complicated and devastating ailment that impacts millions of humans globally, ranking as the second main motive of loss of life (National Cancer Institute, 2021). Despite tremendous improvements in oncology, cancer remains a tremendous public health mission, with increasing instances and deaths projected for the future. Oncology features a multidisciplinary technique related to diverse healthcare experts: clinical oncologists, surgical oncologists, radiation oncologists, pathologists, and radiologists. These professionals collaborate to offer comprehensive care to cancer patients, from analysis to treatment and past (National Cancer Institute, 2021). Cancer is a complicated ailment that may originate in any part of the frame and spread to other regions through the bloodstream or lymphatic system. It requires specialized knowledge and a holistic technique to control and deal with effectively. The causes of most cancers are multifactorial, regarding genetic and environmental factors. Age, genetics, lifestyle selections like tobacco use and harmful diet, exposure to environmental pollutants, and infection with positive viruses like human papillomavirus (HPV) and hepatitis B and C are considered hazard elements for most cancers (American Society of Clinical Oncology, 2021). Despite cancer's challenges, enormous progress has been made in the oncology field in recent years. Technological advancements, precision medication, and immunotherapy have contributed to stepped-forward outcomes for cancer patients (Bouvard et al., 2009). However, there is nevertheless a whole lot of work to be performed, and ongoing research is important to broaden new and greater effective remedies for this devastating sickness. In this paper, we will delve into various elements of oncology, its ancient heritage, distinctive styles of cancer, diagnostic methods employed, and the range of remedy alternatives to be had. By exploring those areas, we propose to gain a comprehensive understanding of oncology and its significance in fighting most cancers.

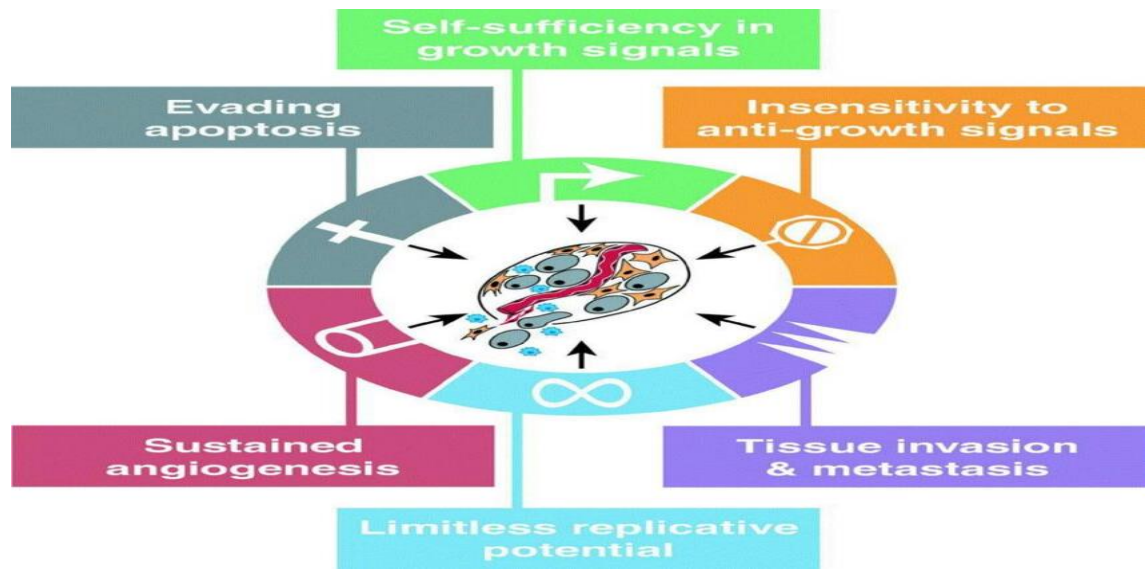

Figure 1. Acquired Capabilities of Cancer (Hanahan & Weinberg, 2000).

**Types of Cancer**

Cancer features an extensive range of sicknesses, each originating from special types of cells in the frame. Some of the most familiar varieties of cancer encompass prostate cancer, breast cancer, lung cancer, and colorectal cancer. Prostate cancer generally affects the prostate gland in men, which is accountable for generating seminal fluid. Breast cancer, then again, develops within the breast tissue and may affect both women and men, even though it is not unusual in women. Lung most cancers originate within the lungs and are frequently associated with tobacco smoking, even though non-people who smoke can also develop this sort of cancer. Colorectal cancer refers to cancer that happens inside the colon or rectum, which might be elements of the digestive machine. Each kind of cancer has distinct traits, including the increased charge, the potential for spreading to other body components (metastasis), and the reaction to particular treatments.

Consequently, treatment tactics for most cancers are tailored to the precise kind and degree of the ailment. This personalized technique ensures that patients acquire the most effective and appropriate treatments, including surgical operation, radiation remedy, chemotherapy, centered therapy, immunotherapy, or a combination of those modalities. Understanding the particular traits of different types of cancers is important for accurate prognosis, powerful remedy planning, and step-forward consequences for sufferers. Ongoing research and improvements in oncology continue to enhance our understanding of those diseases, mainly to the improvement of more targeted and customized healing procedures.

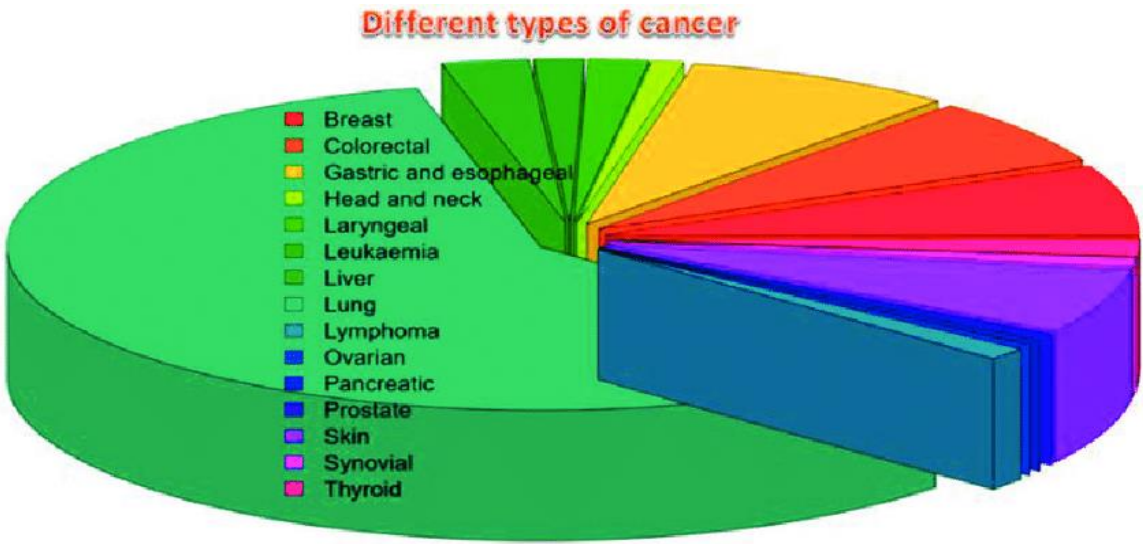

**Diagnostic Methods**

Diagnosing most cancers can indeed be a difficult challenge, and early detection is important for successful remedy results. In the sphere of oncology, diverse diagnostic strategies are employed to discover and verify the presence of most cancers in sufferers. Imaging exams play a considerable position in cancer analysis. X-rays, CT scans (computed tomography), and MRI scans (magnetic resonance imaging) are typically used as imaging techniques. X-rays use radiation to create images of the frame's internal systems, allowing doctors to visualize abnormalities, which include tumors. CT scans offer precise go-sectional snapshots simultaneously as MRI scans and use effective magnets and radio waves to generate notably detailed photographs of the frame's tissues. These imaging checks help identify the location, length, and volume of tumors, aiding in the analysis and staging of cancer. Another essential diagnostic approach is a biopsy. A biopsy includes the removal of a small tissue sample from a suspicious location or tumor, which is then examined beneath a microscope by a pathologist. This examination allows for determining whether the tissue is cancerous and, if so, the type and grade of most cancers. Biopsies can be performed through diverse strategies, such as needle biopsies, where a skinny needle is used to extract cells from the tumor, or surgical biopsies, which involve eliminating a larger tissue pattern. In addition to imaging tests and biopsies, blood exams can also be used to stumble on positive varieties of most cancers. These assessments analyze precise markers or substances in the blood that could indicate the presence of cancer. For instance, prostate-unique antigen (PSA) blood checks are typically used to display prostate cancer. At the same time, positive tumor markers like CA-125 and CEA can be measured to hit ovarian and colorectal cancers, respectively. However, it is crucial to observe that blood tests on my own are not sufficient for a definitive cancer prognosis and are frequently used in conjunction with different diagnostic methods. By utilizing an aggregate of imaging tests, biopsies, and blood exams, healthcare specialists can diagnose most cancers, decide their level and volume, and increase the perfect treatment plan tailored to the person affected. These diagnostic techniques retain to adapt, with ongoing studies and technological improvements contributing to stepped-forward accuracy and performance in most cancer detection.

**Treatment Options**

The treatment of cancer is pretty individualized and relies upon different factors, such as the type of cancer, its stage, the location and length of the tumor, in addition to the patient's age, standard fitness, and private possibilities. Surgery is a not unusual remedy alternative for lots of sorts of cancers. It entails the removal of the cancerous tumor and surrounding tissue. Surgical operation intends to take away the most cancers and save you its spread to other parts of the frame. In some instances, minimally invasive strategies, together with laparoscopic or robotic-assisted surgery, may be used to lessen the invasiveness and restoration time. Radiation therapy, also referred to as radiotherapy, makes use of high-power radiation beams to kill cancer cells or inhibit their growth. It may be brought externally, where the radiation supply is directed at the tumor from outside the body, or internally through brachytherapy, where radioactive implants are located immediately into or near the tumor. Radiation therapy may be used as the primary treatment or in a mixture with other remedies, including surgical operation or chemotherapy.

Chemotherapy involves using effective drugs to kill most cancer cells or inhibit their growth. It can be administered orally, intravenously, or via different techniques. Chemotherapy is regularly used when most cancers have unfolded to distinctive components of the body or in instances where surgical procedures or radiation therapy alone might not be enough. It is occasionally used earlier than surgical operation to decrease tumors (neoadjuvant chemotherapy) or after surgical operation to kill any final most cancer cells (adjuvant chemotherapy). Immunotherapy is a remedy that harnesses the electricity of the frame's immune machine to combat cancer. It works by stimulating the immune device or introducing materials that beautify its potential to apprehend and assault most cancer cells. Immunotherapy can be in the shape of monoclonal antibodies, immune checkpoint inhibitors, or cancer vaccines. This method has shown exceptional fulfillment in treating diverse cancers and has extensively stepped forward effects for a few patients. The targeted remedy is a remedy approach specializing in particular proteins, genes, or other molecules that play a critical role in the growth and spread of cancer cells. Unlike chemotherapy, which influences healthy and cancerous cells, focused remedy capsules selectively target cancer cells, minimizing damage to wholesome tissues. These pills can block the indicators that promote most cancer cell growth or deliver toxic materials directly to cancer cells. Targeted therapy is regularly used while unique genetic mutations or alterations are present inside the cancer cells. Hormone therapy is mostly used for hormone-sensitive cancers, together with breast and prostate cancer. It works by blocking off the effects of positive hormones or decreasing their manufacturing within the frame. For example, hormone remedies for breast cancer might also involve the use of drugs that block estrogen receptors, thereby stopping estrogen from stimulating the growth of cancer cells.

In many cases, an aggregate of treatments can be used. This approach is referred to as a multimodal or combination remedy and aims to target most cancers from one-of-a-kind angles, increasing the probability of a hit remedy. The specific mixture of treatments depends on the person affected person's desires and is decided with the aid of a multidisciplinary team of healthcare professionals. It's important to note that remedy plans constantly evolve, and new remedy options are being developed via ongoing research and scientific trials. The desire for treatment is a collaborative selection among the affected person and their healthcare team, taking into account the ability advantages, risks, and facet consequences of each treatment alternative.

**Advances in Oncology**

In recent years, oncology has witnessed notable advancements, mainly to improved effects and excellent life for patients with most cancers. One large improvement is the emergence of precision medicinal drugs as a promising approach to most cancer remedies. Precision remedy involves using genetic testing to identify particular mutations or biomarkers in most cancer cells. With expertise in the specific genetic makeup of a patient's cancer, medical doctors can tailor remedy plans to target these unique mutations or biomarkers. This personalized method has proven wonderful potential in improving remedy efficacy and lowering the side outcomes of traditional treatment options. Immunotherapy is another exciting area of research in oncology. This method harnesses the strength of the immune device to combat most cancers. Immunotherapy tablets work by stimulating the immune machine or blocking off immune checkpoints, which might be proteins that prevent immune cells from attacking cancer cells. This treatment modality has tested tremendous achievement in several kinds of cancers, which include cancer and lung cancer. It has shown long-lasting responses or even led to long-term remission in some instances, presenting a new desire for sufferers with superior or metastatic cancers. Advances in technology have also revolutionized the field of oncology. Robotic surgical operation, for instance, has changed the way complicated surgical strategies are carried out. With the assistance of robotic structures, surgeons can perform with stronger precision, dexterity, and visualization. This minimally invasive method permits smaller incisions, decreased blood loss, and faster healing instances for patients. Robotic surgical treatment has been successfully utilized in numerous cancer surgeries, such as prostate, gynecological, and gastrointestinal processes.

Furthermore, technological advancements have significantly contributed to the sector of clinical imaging, allowing greater correct and specific diagnoses. High-decision imaging techniques, along with positron emission tomography (PET), magnetic resonance imaging (MRI), and computed tomography (CT), provide unique information about the area, length, and quantity of tumors. These superior imaging technologies assist in guiding remedy planning, screening remedy response, and coming across cancer recurrence at an advanced stage. In addition to precision medicine, immunotherapy, and technological improvements, ongoing research continues to explore different revolutionary treatment modalities, including gene therapy, CAR-T mobile remedy, and nanotechnology-primarily based drug delivery systems. These current techniques maintain great promise and may also revolutionize cancer treatment. It is essential to be aware that while full-size progress has been made, there are nonetheless challenges to triumph over inside the oncology discipline. The development of resistance to treatments, the complexity of tumor heterogeneity, and the want for powerful techniques to save cancer are regions that require similar exploration and research. The mixture of precision medicinal drugs, immunotherapy, technological improvements, and ongoing research has introduced a new era of wish and optimism in combating most cancers. These improvements continue to form the oncology landscape, supplying sufferers with improved treatment alternatives, higher outcomes, and more desirable pleasant lifestyles.

**Challenges in Oncology**

Despite the vast development carried out in the discipline of oncology, there remain numerous challenges that require interest and backbone. One of the most pressing challenges is the excessive fee of most cancer remedies. Many present-day treatments and focused capsules are prohibitively high-priced, rendering them inaccessible to most of the populace. This economic burden creates disparities in cancer care, with some patients not able to have the funds for the necessary treatments, leading to suboptimal outcomes. Addressing the difficulty of affordability and ensuring equitable access to the most effective cancer treatments is crucial so that you can provide exceptional care for all sufferers. The principal task in most cancer treatments is the development of drug resistance. Cancer cells can evolve and adapt, mainly to resistance against chemotherapy and traditional treatments. This resistance can result in remedy failure and avert long-term remission. Researchers are actively investigating the mechanisms behind drug resistance and operating to increase progressive remedy strategies to bypass it. This consists of identifying new drug objectives, growing aggregate treatment options, and utilizing immunotherapy approaches to triumph over resistance mechanisms hired by way of cancer cells. Furthermore, the complexity and heterogeneity of tumors present an ongoing mission in oncology. Each tumor and numerous mobile populations with varying genetic and molecular traits are particular. This heterogeneity contributes to variations in treatment response and ailment development amongst sufferers. Researchers are striving to understand tumor biology better and expand customized treatment methods that consider the specific features of each affected person's tumor. This includes the usage of superior genomic profiling, liquid biopsies, and synthetic intelligence techniques to pick out tumor subtypes and expect remedy responses extra correctly.

Additionally, the prevention of cancer remains an important venture in oncology. While advancements in early detection and screening applications have helped pick out cancer at earlier levels, preventing the disorder from going on within the first area is of extreme importance. Efforts are being made to promote wholesome lifestyles, raise focus on chance factors, and expand powerful preventive techniques, along with vaccines for certain varieties of cancer (e.g., HPV vaccine for cervical cancer prevention). By specializing in prevention, the burden of cancer can be drastically decreased, leading to improved public fitness effects. In the end, while great development has been made inside the oncology discipline, there are still vast demanding situations to address. These challenges include the excessive price of most cancer treatments, the improvement of drug resistance, the complexity of tumor heterogeneity, and the want for effective cancer prevention strategies. Continued studies, collaboration, and innovation are important in overcoming those demanding situations and enhancing consequences for patients with cancer. By addressing these limitations, we can paint closer to a future wherein cancer treatment is offered, powerful, and personalized, ultimately accomplishing better outcomes and enhancing the lives of patients worldwide.

**Conclusion**

As a crucial field of medicine, oncology is pivotal in preventing, analyzing, and treating cancer. Despite the sizable progress performed in this area, most cancers remain a formidable public health undertaking, necessitating ongoing studies to expand novel and more powerful remedies for this devastating disorder. Advances in precision remedies, immunotherapy, and technology have revolutionized cancer care, leading to progressed results for patients. Precision medication permits centered treatments primarily based on an affected person's specific genetic profile, growing treatment efficacy and minimizing aspect effects. Immunotherapy, then again, harnesses the strength of the immune machine to understand and smash most cancer cells, providing new hope for sufferers with formerly restricted treatment alternatives. Furthermore, improvements in generation have facilitated early detection through progressed imaging strategies and more desirable surgical precision. However, regardless of those advancements, challenges persist within the oncology discipline. The high cost of cancer remedies poses a great barrier, restricting the right of entry to the best care for many sufferers. Affordability and equitable entry to progressive remedies are vital for ensuring that every individual can benefit from the latest healing interventions. Efforts to lessen the financial burden of cancer care and explore alternative funding fashions are critical to address this task. Another bold obstacle is the improvement of drug resistance. Cancer cells can adapt and emerge as resistant to traditional treatment plans, rendering them ineffective and impeding lengthy-term remission. Researchers are diligently investigating the underlying mechanisms of drug resistance to develop strategies that can triumph over or skip this task. Combination therapies, targeted cures, and immunotherapies tailored to combat drug-resistant cancer cells are being explored to enhance treatment outcomes and prolong survival. In addition to remedy, emphasis is likewise on most cancer prevention and early detection. Public health campaigns are instrumental in raising awareness of the significance of a wholesome way of life picks, including tobacco cessation, a balanced weight loss program, ordinary workouts, and cancer screening programs. Early detection of cancer notably improves prognosis and treatment consequences, as it lets in for well-timed intervention while the disorder remains localized and more amenable to healing treatment alternatives. Ultimately, the goal of oncology is to provide first-class feasible consequences for sufferers with cancer, whether via healing or palliative remedy strategies. By collaborating across disciplines and constantly advancing our information on this complex disorder, we can make vast strides in improving the lives of thousands and thousands of people stricken by cancer worldwide. Through research-driven improvements, personalized care, and holistic help, we will beautify affected person outcomes, sell survivorship, and alleviate the physical, emotional, and socioeconomic burdens associated with cancer. Together, we can work closer to a future where cancer is preventable, detectable at its earliest levels, and effectively controlled, making sure of a better nice of life for all those impacted using this ailment.